\documentclass[runningheads]{llncs}
\usepackage{graphicx}
\usepackage{makecell}
\usepackage[multiple]{footmisc}
\usepackage{here}
\usepackage{marvosym}
\usepackage{hyperref}
\usepackage{subcaption}
\begin{document}
\titlerunning{Segmentation, Classification, and Quality Assessment for OCTA of DR}
\title{Segmentation, Classification, and Quality Assessment of UW-OCTA Images for the Diagnosis of Diabetic Retinopathy \thanks{This work is supported by the ANR RHU project Evired. This work benefits from State aid managed by the French National Research Agency under ``Investissement d'Avenir'' program bearing the reference ANR-18-RHUS-0008.}}
%
%


\author{Yihao Li\inst{1,2}\textsuperscript{(\Letter)}
\and
Rachid Zeghlache\inst{1,2}
\and
Ikram Brahim\inst{1,2}
\and
Hui Xu\inst{1}
\and
Yubo Tan\inst{3}
\and
Pierre-Henri Conze\inst{1,4} 
\and
Mathieu Lamard\inst{1,2} 
\and
Gwenolé Quellec\inst{1} 
\and
Mostafa El Habib Daho \inst{1,2}\textsuperscript{(\Letter)}
}

\authorrunning{Y. Li et al.}
%
\institute{
LaTIM UMR 1101, Inserm, Brest, France \and
University of Western Brittany, Brest, France \and
University of Electronic Science and Technology of China, Chengdu, China\and
IMT Atlantique, Brest, France \\
\email{Yihao.Li@etudiant.univ-brest.fr}\\
\email{mostafa.elhabibdaho@univ-brest.fr}
}
\maketitle              

\begin{abstract}
Diabetic Retinopathy (DR) is a severe complication of diabetes that can cause blindness. Although effective treatments exist (notably laser) to slow the progression of the disease and prevent blindness, the best treatment remains prevention through regular check-ups (at least once a year) with an ophthalmologist. 
Optical Coherence Tomography Angiography (OCTA) allows for the visualization of the retinal vascularization,
and the choroid at the microvascular level in great detail. This allows doctors to diagnose DR with more precision. 
In recent years, algorithms for DR diagnosis have emerged along with the development of deep learning and the improvement of computer hardware. However, these usually focus on retina photography. There are no current methods that can automatically analyze DR using Ultra-Wide OCTA (UW-OCTA). 
The Diabetic Retinopathy Analysis Challenge 2022 (DRAC22) provides a standardized UW-OCTA dataset to train and test the effectiveness of various algorithms on three tasks: lesions segmentation, quality assessment, and DR grading. 
In this paper, we will present our solutions for the three tasks of the DRAC22 challenge. The obtained results are promising and have allowed us to position ourselves in the TOP 5 of the segmentation task, the TOP 4 of the quality assessment task, and the TOP 3 of the DR grading task. The code is available at \url{https://github.com/Mostafa-EHD/Diabetic_Retinopathy_OCTA}.

\end{abstract}
\section{Introduction}
Diabetes, specifically uncontrolled diabetes causing Diabetic Retinopathy (DR), is among the leading causes of blindness. DR is a condition that affects approximately 78\% of people with a history of diabetes of 15 years or more \cite{https://doi.org/10.1111/aos.14299}. In the early stages, DR is considered a silent disease. For this reason, seeing an ophthalmologist regularly, especially if you have diabetes, is essential to avoid the risk of serious complications, including blindness.\\

DR is diagnosed by visually inspecting fundus images for retinal lesions such as Microaneurysms (MA), Intraretinal Microvascular Anomalies (IRMA), areas of Non-Perfusion, and Neovascularization. Fundus photography and fundus fluorescein angiography (FFA) are the two most commonly used tools for DR screening. Traditional diagnosis of DR relies mainly on these two modalities, especially for Proliferative Diabetic Retinopathy (PDR), which seriously endangers visual health. However, fundus photography has difficulty detecting early or small neovascular lesions, and FFA is an invasive fundus imaging that cannot be used in allergic patients, pregnant women, or those with impaired liver and kidney function.

Optical Coherence Tomography Angiography (OCTA) is a new non-invasive imaging technique that generates volumetric angiographic images in seconds. It can display both structural, and blood flow information \cite{deCarlo2015}. Due to the quantity and quality of the information provided by this modality, OCTA is being increasingly used for diagnosing DR at the early stages. In addition, the Swept-Source OCTA (SS-OCTA) allows the individual assessment of choroidal vascularization and the Ultra-Wide Optical Coherence Tomography (UW-OCTA) imaging modality has shown a more significant pathological burden in the retinal periphery that was not captured by typical OCTA \cite{QIMS21249}.\\

Several DR diagnosis algorithms have emerged in recent years through improved computer hardware, deep learning, and data availability \cite{Quellec17}\cite{ExplAIn}\cite{Zeghlache}\cite{Yihao}\cite{Atwany} \cite{00f07def2e1c497d99207b316b574b21}\cite{Quellec22}\cite{10.3389/fpubh.2022.971943}\cite{LIU2022100512}. Some works have already used SS-OCTA to assess the qualitative characteristics of DR \cite{Schaal} and others have used UW-OCTA on DR analysis \cite{QIMS21249}\cite{Russell}\cite{Yihao}. However, there is currently no work that can automatically analyze DR using UW-OCTA. In the DR analysis process, the image quality of the UW-OCTA must first be evaluated, and the best quality images are selected. Then, DR analysis, such as lesion segmentation and PDR detection, is performed. Therefore, it is crucial to build a full pipeline to perform automatic image quality assessment, lesion segmentation, and PDR detection. \\

The Diabetic Retinopathy Analysis Challenge 2022 (DRAC22) provides a standardized UW-OCTA dataset to train and test the effectiveness of various algorithms. \\
DRAC22 is a first edition associated with MICCAI 2022 that offers three tasks to choose from:
\begin{itemize}
    \item Task 1: Segmentation of DR lesions.
    \item Task 2: Image quality assessment.
    \item Task 3: Classification of DR.
\end{itemize}

This article will present our three proposed solutions to solve each task of the DRAC22 challenge.

\section{Materials and methods}

\subsection{Datasets}
The instrument used to gather the dataset in this challenge was an SS-OCTA system (VG200D, SVision Imaging, Ltd., Luoyang, Henan, China), which works near 1050nm and features a combination of industry-leading specifications, including an ultrafast scan speed of 200,000 AScans per second \footnote[1]{\url{https://drac22.grand-challenge.org/Data/}}\footnote[2]{\url{https://svisionimaging.com/index.php/en\_us/home/}}.\\

The following table summarizes the data collected by the DRAC22 Challenge organizers. All images are 2D en-face images.

\begin{table}[H]
\label{tab2}
\centering
\caption{DRAC 2022 datasets}
\label{tab:my-table}
\begin{tabular}{|l|c|c|}
\hline
\multicolumn{1}{|c|}{Task}   & \# Training images & \# Test images \\ \hline
Task 1 - Segmentation & 109                & 65             \\ \hline
Task 2 - Quality assessment & 665                & 438            \\ \hline
Task 3 - Classification & 611                & 386            \\ \hline
\end{tabular}
\end{table}

\begin{figure}
\centering
\begin{subfigure}[b]{.75\linewidth}
\includegraphics[width=\linewidth]{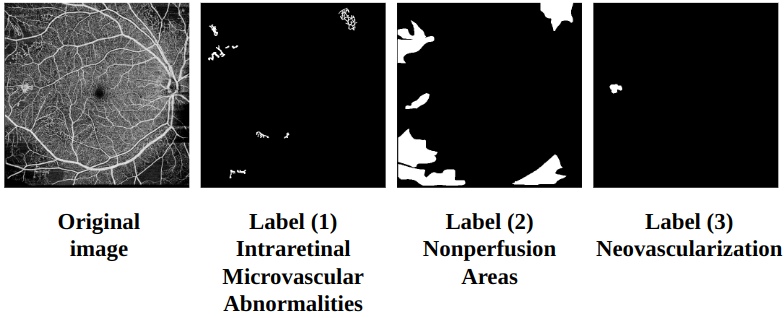}
\caption{Segmentation dataset}\label{fig:seg}
\end{subfigure}

\begin{subfigure}[b]{.49\linewidth}
\includegraphics[width=\linewidth]{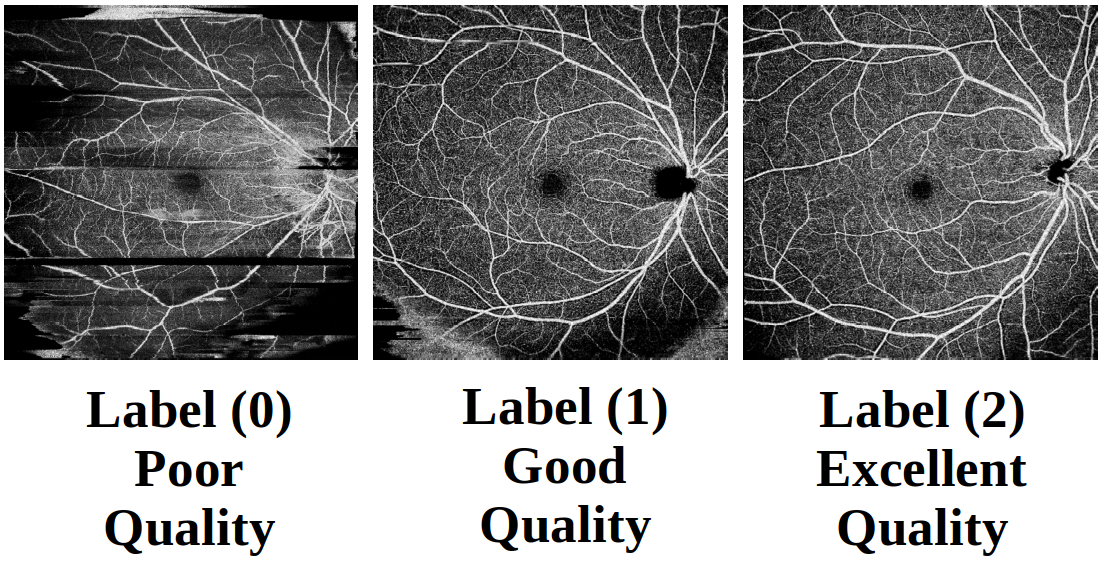}
\caption{Quality assessment dataset}\label{fig:qual}
\end{subfigure}
\begin{subfigure}[b]{.49\linewidth}
\includegraphics[width=\linewidth]{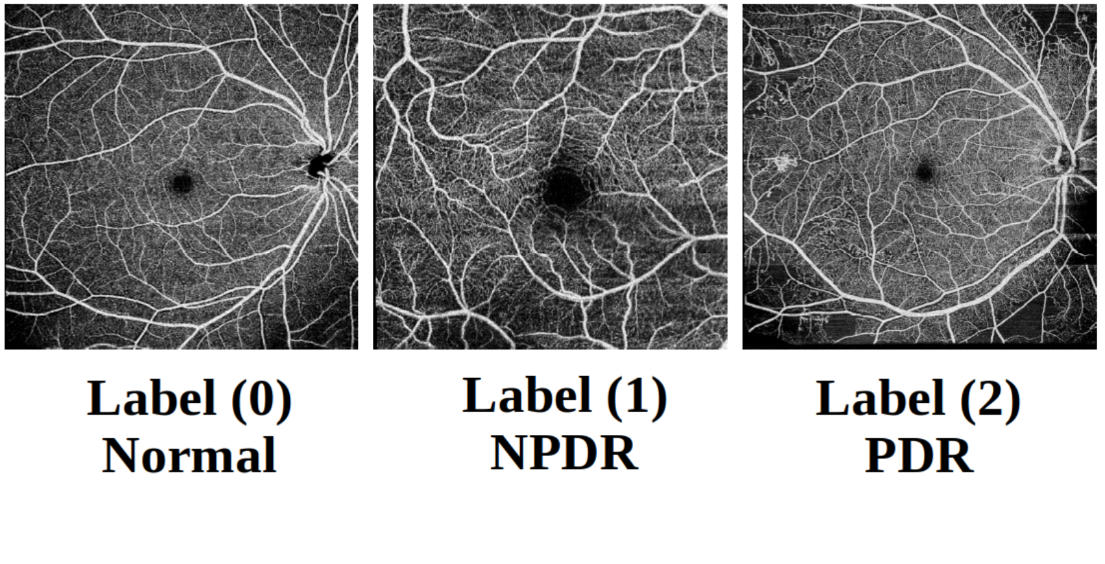}
\caption{Diabetic Retinopathy grading dataset}\label{fig:class}
\end{subfigure}
\caption{DRAC22 Challenge Dataset}
\label{fig:drac22}
\end{figure}

The training set consists of 109 images and corresponding labels for the first task. The dataset, as shown in Figure \ref{fig:seg}, contains three different Diabetic Retinopathy Lesions: Intraretinal Microvascular Abnormalities (1), Nonperfusion Areas (2), and Neovascularization (3). The test set consists of 65 images.\\

For the second task, quality assessment, the organizers propose a dataset of 665 and 438 images for training and testing, respectively. These images (see Figure \ref{fig:qual}) are grouped into three categories: Poor quality level (0), Good quality level (1), and Excellent quality level (2). \\

The third dataset is dedicated to the classification task. It contains 611 images for learning and 386 for testing, grouped into three different diabetic retinopathy grades as shown in Figure \ref{fig:class}: Normal (0), NPDR (1), and PDR (2).\\

\subsection{Task 1 - Segmentation}

In this section, we introduce the models and techniques used to solve the segmentation problem: nnU-Net and V-Net.  \\

U-Net is a simple and successful architecture that has quickly become a standard for medical image segmentation \cite{ronneberger2015unet}. However, adapting U-Net to new problems is not straightforward, as the exact architecture and the different parameters would have to be chosen. The no-new-U-Net (nnU-Net) method provides an automated end-to-end pipeline that can be trained and inferred on any medical dataset for segmentation \cite{https://doi.org/10.48550/arxiv.1809.10486}. nnU-Net outperformed state-of-the-art architectures in the Medical Decathlon Challenge\footnote[3]{\url{http://medicaldecathlon.com/results/}}.    \\

After analyzing the dataset of the segmentation task, we noticed that the three labels overlapped on several images, so we opted to train three binary segmentation models for the segmentation of each of the labels. \\

We first trained an nnU-Net model for each label. Initial results showed that the trained nnU-Net for label 2 (nonperfusion areas) performed well. However, the other two models trained on label 1 (intraretinal microvascular anomalies) and label 3 (neovascularization) did not learn well, and the results were poor. \\
As a second solution, we trained a binary V-Net model for the segmentation of each of the labels: 1 and 3. V-Net is a U-Net-based network that incorporates residual blocks into the network. Residual linking encourages the training process to converge faster \cite{milletari2016vnet}. The hyperparameters can be continuously tested to improve segmentation performance over nnU-Net. \\

We observed that the model for label 3 tended to over-segment. To alleviate this issue, we added a classification step to predict the probability that an image contains label 3. The added classifier (based on ResNet) allowed us to improve our results on the test base since an image that has been classified as not containing label 3 will not be segmented. The models and parameters are summarized in Table~\ref{tab1}. \\

\begin{table}[htb]
\centering
\caption{Implementations of different labels}\label{tab1}
\resizebox{\textwidth}{35mm}{
\begin{tabular}{|l|c|c|c|c|c|}
\hline
\makecell[c]{Segmentation \\Task}   & Architecture&  Image Size & Data Augmentation& Loss & Optimizer\\ \hline
\makecell[c]{Label 1 - \\Intraretinal \\ Microvascular \\ Abnormalities} & 2D V-Net   & \makecell[c]{1024$\times$1024} & \makecell[c]{RandomAffine \\ Rand2DElastic \\ Default Data \\Augmentation  }&\makecell[c]{Dice loss} &    \makecell[c]{Adam \\ lr = 1e-3  \\ ExponentialLR \\(gamma = 0.99)}     \\ \hline
\makecell[c]{Label 2 - \\ Nonperfusion \\Areas} & 2D nnU-Net   & \makecell[c]{1024$\times$1024\\Normalization} & \makecell[c]{Random rotations, \\Random scaling, \\Random elastic \\Deformations, \\Gamma correction \\ and mirroring }  &   \makecell[c]{Dice loss + \\Cross-entropy \\loss}&  \makecell[c]{Adam \\ lr = 0.01 }  \\ \hline
\makecell[c]{Label 3 - \\  Neovascularization} & 2D V-Net  & \makecell[c]{1024$\times$1024} & \makecell[c]{RandomAffine \\ Rand2DElastic \\ Default Data \\Augmentation  }&\makecell[c]{Dice loss} &    \makecell[c]{Adam \\ lr = 1e-3  \\ ExponentialLR \\(gamma = 0.99)}     \\ \hline
\makecell[c]{Label 3 \\ Classifier} & ResNet101  & \makecell[c]{1024$\times$1024} & \makecell[c]{ Default Data \\Augmentation  }&\makecell[c]{Cross-entropy \\loss} &    \makecell[c]{Adam \\ lr = 1e-4\\ weight\_decay=1e-4  \\ ExponentialLR \\(gamma = 0.99)}     \\ \hline

\end{tabular}}
\end{table}

Dice and cross-entropy loss are used to train the nnU-Net network. The dice loss formulation is adapted from the variant proposed in \cite{https://doi.org/10.48550/arxiv.1809.10486}. It is implemented as follows:

\begin{equation}
L_{dice} = - \frac{2}{\left | K \right | } \sum_{k\in K}^{} \frac{ {\textstyle \sum_{i\in I}^{}} u_{i}^{k}v_{i}^{k}  }{ {\textstyle \sum_{i\in I}^{}} u_{i}^{k}+  {\textstyle \sum_{i\in I}^{}} v_{i}^{k}   } 
\end{equation}

Where $u$ is the softmax output of the network and $v$ is the one hot encoding for the ground truth segmentation map. Both $u$ and $v$ have shape $I \times K $  with $i \in I$ being the number of pixels in the training patch/batch and $k \in K$ being the classes.\\

Adam optimizer is used to train the nnU-Net network with an initial learning rate of 0.01. A five-fold cross-validation procedure is used, and the model has been trained over 1000 epochs per fold with a batch size is fixed to 2. Whenever the exponential moving average of the training losses did not improve by at least $ 5 \times 10^{-3}$ within the last 30 epochs, the learning rate was reduced by factor 5. The training was stopped automatically if the exponential moving average of the validation losses did not improve by more than $ 5 \times 10^{-3}$ within the last 60 epochs, but not before the learning rate was smaller than $ 10^{-6}$ \cite{https://doi.org/10.48550/arxiv.1809.10486}.\\

Dice loss (include\_background = False) is used to train the V-Net network. Besides the default data augmentation (random crop, random flip, and random rotation), RandomAffine and Rand2DElastic were also used. And mean\_Dice is used to select the best checkpoint. The training epoch is 1000, the optimzer is Adam, and the batch size is 3. Finally, for the classifier of label 3, the batch size is 4 and the epoch is 500.

\subsection{Task 2\&3 - Quality assessment \& Classification of DR}

As both Task 2 and Task 3 involved three-labels classifications, the pipeline was the same. To verify the performance of the different models, we used five-fold cross-validation to test six architectures (17 backbones): ResNet \cite{https://doi.org/10.48550/arxiv.1512.03385}, DenseNet \cite{https://doi.org/10.48550/arxiv.1608.06993}, EfficientNet \cite{https://doi.org/10.48550/arxiv.1905.11946}, VGG \cite{https://doi.org/10.48550/arxiv.1409.1556}, ConvNeXt \cite{https://doi.org/10.48550/arxiv.2201.03545}, Swin-Transformer \cite{https://doi.org/10.48550/arxiv.2103.14030}. \\
These backbones were pre-trained using ImageNet and originated from the timm library \cite{rw2019timm}. We kept the original size of the image during training, i.e., 1024$\times$1024.

We have performed several experiments to improve our models using different optimizers (SGD, Adam, AdamW, RMSprop), schedulers (ReduceLROnPlateau, ExponentialLR, LambdaLR, OneCycleLR), and loss functions. Among the strategies we have tested:

\begin{enumerate}
    \item Training with the \textbf{Cross Entropy (CE)}
    \item Training with the \textbf{Weighted Cross Entropy (WCE)}
    \item Training with the $\mathbf{KappaLoss}$
    \item Training with the $\mathbf{WCE}$ + $\mathbf{\lambda\cdot KappaLoss}$
    \item Training with the $\mathbf{\alpha \cdot}$$\mathbf{WCE}$ + $(1-\alpha)\cdot$$\mathbf{KappaLoss}$
    \item Training with the $\mathbf{WCE}$ and finetuning with the $\mathbf{KappaLoss}$
    
\end{enumerate}

With: 
\begin{itemize}
    \item $KappaLoss$ = 1 - Quadratic Weighted Kappa Score. The Quadratic Weighted Kappa Score was computed using the TorchMetrics framework.
    \item $\lambda$ is a balancing weight factor and $\alpha$ is a weight factor that decreases linearly with the number of epochs.
\end{itemize}

We faced convergence issue during the training of our approaches with the KappaLoss alone or in combination with WCE in our objective function. We were not able to optimize the weights of our models correctly. The best results were obtained with the CE.  \\

CE is used as a loss function, and Kappa is used to select the best checkpoint. The default data augmentation and Adam optimizer with an initial learning rate of $ 10^{-4}$ (weight\_decay=$ 10^{-4}$) were used to train different backbones. The learning rate decay strategy is ExponentialLR with gamma equal to 0.99. The training epoch is 1000 and the batch size is 4.\\

\begin{figure}[!htb]
\centering
\includegraphics[width=1.05\textwidth]{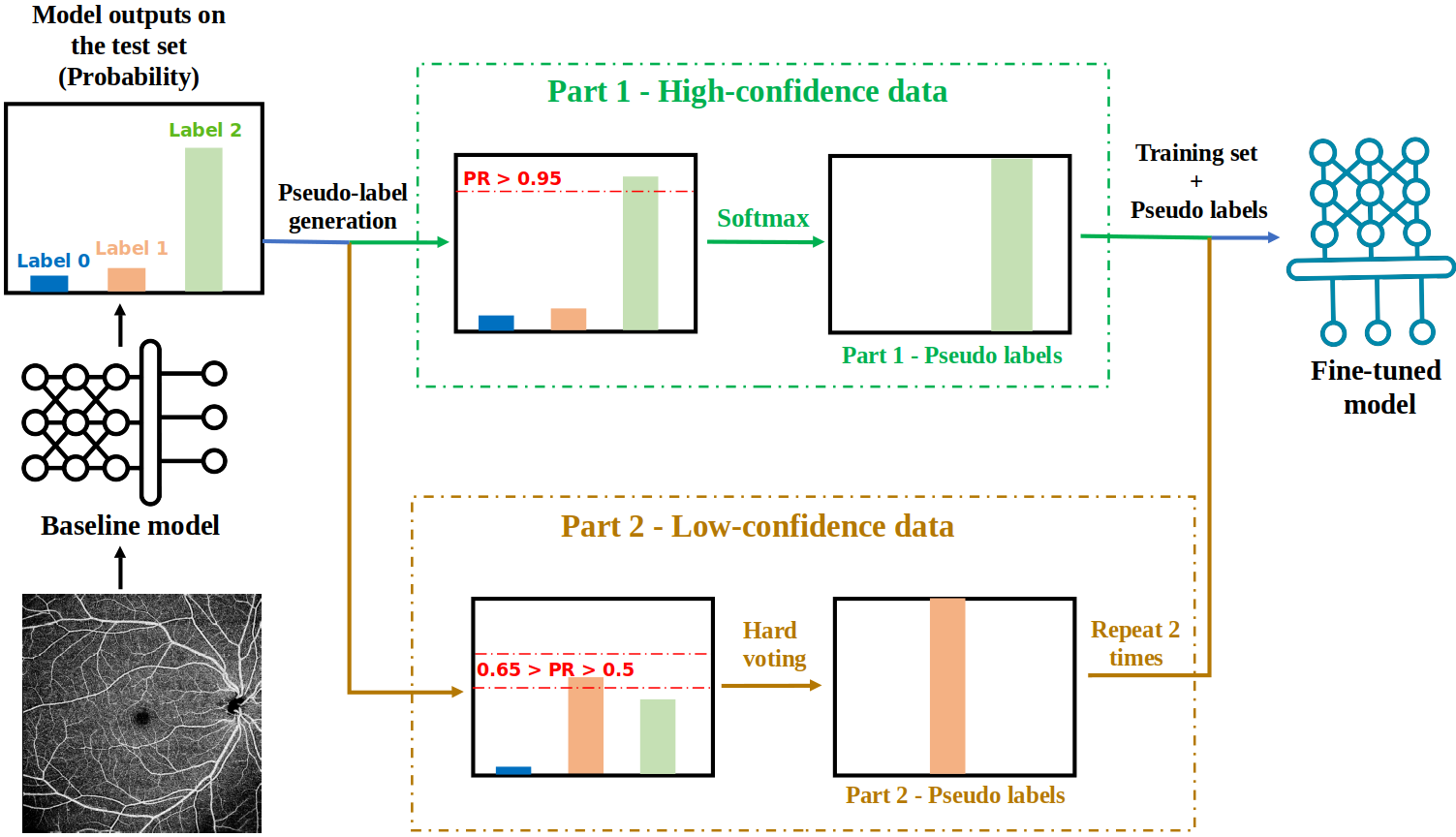}
\caption{Our pseudo-label learning method.} \label{fig1}
\end{figure}

Once our baselines were trained, we proceeded to improve them. We used the pseudo-labeling technique. Pseudo-labeling is a process that consists in adding confidently predicted test data to the training data and retraining the models \cite{lee2013pseudo}.
Through pseudo-label learning, we improved the classification performance of the model after obtaining a good baseline model. Fig.~\ref{fig1} illustrates our proposed pseudo-label learning method.

According to the probabilities generated by the baseline model on the test set, we have separated the result into two chunks: the first is the high confidence data, and the second is the low confidence data. As with traditional pseudo-label learning methods \cite{lee2013pseudo}, we considered data with predicted classification probabilities greater than 0.95 to be high-confidence and passed their probability through the Softmax function to determine the pseudo-label. \\

Repeated training on hard-to-classify samples, like those involved in the Online Hard Example Mining method \cite{shrivastava2016training}, can enhance the model's performance. Therefore, we used data with probabilities between 0.5 and 0.65, and generated pseudo-labels based on five backbones through hard voting. In order to ensure the accuracy of pseudo-labeling of low-confidence data, the following filtering rules were applied.\\
\begin{enumerate}
    \item The baseline model results as pseudo-labels if at least two of the other four backbones have the same result.
    \item The results of the other four backbones as pseudo-labels if they are consistent.
\end{enumerate}
Those remaining cases cannot be pseudo-labeled, so their data will not be used.

Since our baseline model performed well, the high-confidence pseudo-labels are more accurate, adding additional data to the training set can improve model robustness. \\
By hard voting and developing filtering rules, we made the pseudo-labels as accurate as possible for low-confidence data. With the help of this part of the data, the model can make more accurate judgments on data without distinctive features (with probabilities of between 0.65 and 0.9) and improve confidence levels on the remaining uncertain data. \\
Both pseudo-labels were added to the training set, and the baseline model was fine-tuned. Due to the small size of the second part of the pseudo-label data, we repeated it twice. Furthermore, the results can be further enhanced by iterative pseudo-label learning.\\

\section{Results and discussion}

\subsection{Evaluation metrics}
For the first task (segmentation), the Dice similarity coefficient (DSC) and the intersection of union (IoU) are used to evaluate the performance of the segmentation methods. \\

The Dice coefficient (also known as the Sørensen–Dice coefficient and F1 score) is defined as two times the area of the intersection of A and B, divided by the sum of the areas of A and B. The IOU (Intersection Over Union, also known as the Jaccard Index) is defined as the intersection's area divided by the union's area.\\

For tasks 2 (quality assessment) and 3 (classification), the organizers propose to use the quadratic weighted kappa and the area under the curve (AUC) to evaluate the performance of classification methods. In particular, they used the macro-AUC One VS One to calculate the AUC value.

\subsection{Segmentation results}

\begin{table}[htb]

\centering
\caption{Segmentation results}
\label{tab3}
\begin{tabular}{|l|c|c|c|}
\hline
\multicolumn{1}{|c|}{Version}   & Label 1 & Label 2 & Label 3\\ \hline
V1 - nnU-Net &  0.2278  & 0.6515 &  0.4621            \\ \hline
V2 - V-Net & 0.4079 & 0.6515 & 0.5259            \\ \hline
V3 - V-Net + Classifier &  0.4079   & 0.6515 & 0.5566            \\ \hline
\end{tabular}
\end{table}
The segmentation task dataset contains 109 images, and each image can have one or more labels. In the training set, there are 86 images that contain the label (1), 106 images that possess the label (2), and only 35 images with the label (3). \\

We first used the nnU-net method to test the segmentation of the three labels. By analyzing the first results, we noticed that the amount of data for the label (2) is relatively large, so nnU-Net obtained good performances (Dice = 0.6515). However, for labels (1) and (3), the model performed poorly. Therefore, we chose V-Net as the backbone of our binary segmentation for both these labels.  \\

We fine-tuned the V-Net by modifying the loss function, the optimizer, the scheduler, the learning rate, etc. During these experiments, we divided the training dataset into two subsets: 90\% for training and 10\% for validation. After several runs, we got good results; label (1) has a dice of 0.4079.\\

After the first tests, we noticed that the distribution of the 65 images in the test set differed from the training set; more than 40 images in the V-Net prediction result contained the label (3). Therefore, we improved our proposed solution by adding a classifier that can detect whether the image contains the label (3) or not. \\

After the training, the model classified 28 images in class (1), and the rest in class (0), which means that only 28 images contain the label (3), and the others are over-segmented. The label (3) was removed from all the images classified as (0) by the counter-classifier. This step improved the segmentation results for label (3) from 0.5259 to 0.5566. \\

\subsection{Quality assessment results}
The second task of the DRAC22 challenge concerns the image quality assessment. The training set consists of 665 data, distributed as follows: 97 images belonging to class (0) Poor quality level, 518 belonging to class (1) Good quality level, and 50 images belonging to class (2) Excellent quality level. The first observation is that the data set is very unbalanced.\\

We used a nested five-fold cross-validation strategy to evaluate different models and backbones. We respected the distribution of the training set in the generation of the folds. For each split, we used four folds for training and validation (20\% random as the validation set and 80\% as the training set) and one fold for testing. A suitable checkpoint is selected from the validation set, and the final performance of the model is calculated using the test set (one-fold data). This strategy avoids overfitting and provides a more accurate representation of the model's classification performance. \\

Table~\ref{label4} summarizes the different backbones used and the results obtained. The model performed poorly on the Fold 4 dataset. On the other hand, the different backbones perform well on Folds 0, 1, 2, and 3. Therefore, we chose the two most optimally performing checkpoints for each fold and tested them on the test set. Their results are shown in Table~\ref{label5}. Val Kappa refers to the one-fold test results in Table 4, whereas Test Kappa refers to the DRAC22 test dataset results. The V5 - VGG19-Fold2 and V2 - VGG16-Fold0 checkpoints performed the best out of the eight selected checkpoints. In order to optimize the use of the training data, we selected the checkpoints V5 - VGG19-Fold2 and performed fine-tuning (random 20\% validation set) on the entire training dataset, which gave us the baseline model V9 - VGG19-Finetune that has a kappa value of 0.7447 on the test set.

\begin{table}[htb]
\centering
\caption{Kappa results for different backbones on a one-fold test set.}
\label{label4}
\begin{tabular}{|l|l|l|l|l|l|l|}
\hline
Backbone                                 & Fold 0 & Fold 1 & Fold 2 & Fold 3 & Fold 4 & Mean   \\ \hline
Resnet50                         & 0.7179 & 0.8247 & 0.7502 & 0.8409 & 0.5585 & 0.7384 \\ \hline
Resnet101                        & 0.7540 & 0.7857 & 0.7515 & 0.8060 & 0.6199 & 0.7434 \\ \hline
Resnet152                        & 0.8135 & 0.7600 & 0.8537 & 0.8559 & 0.5291 & 0.7624 \\ \hline
Resnet200d                       & 0.5488 & 0.7910 & 0.7889 & 0.8593 & 0.5486 & 0.7073 \\ \hline
Densenet121                      & 0.8525 & 0.7814 & 0.8299 & 0.7966 & 0.5274 & 0.7575 \\ \hline
Densenet161                      & 0.8357 & 0.7666 & 0.8761 & 0.8358 & 0.5569 & 0.7742 \\ \hline
Densenet169                      & 0.7942 & 0.7547 & 0.8680 & 0.7908 & 0.4468 & 0.7309 \\ \hline
Densenet201                      & 0.7980 & 0.8028 & 0.8289 & 0.8563 & 0.5132 & 0.7598 \\ \hline
Efficientnet\_b0                 & 0.7048 & 0.7617 & 0.7859 & 0.8246 & 0.5367 & 0.7227 \\ \hline
Efficientnet\_b1                 & 0.8267 & 0.7503 & 0.8349 & 0.7958 & 0.5759 & 0.7567 \\ \hline
Efficientnet\_b2                 & 0.8406 & 0.8039 & 0.8434 & 0.8563 & 0.5311 & 0.7750 \\ \hline
Efficientnet\_b3                 & 0.7710 & 0.7787 & 0.8821 & 0.7768 & 0.4801 & 0.7377 \\ \hline
Efficientnet\_b4                 & 0.8367 & 0.8202 & 0.8855 & 0.8468 & 0.5254 & 0.7829 \\ \hline
VGG11                            & 0.8420 & 0.7730 & 0.8795 & 0.8606 & 0.6199 & 0.7950 \\ \hline
VGG16                            & 0.8592 & 0.8231 & 0.8551 & 0.9168 & 0.5077 & 0.7923 \\ \hline
VGG19                            & 0.8787 & 0.8042 & 0.8933 & 0.8632 & 0.3192 & 0.7517 \\ \hline
VGG13                            & 0.8302 & 0.8409 & 0.8504 & 0.8233 & 0.6704 & 0.8030 \\ \hline
Convnext\_base                   & 0.8343 & 0.8118 & 0.8531 & 0.8516 & 0.5451 & 0.7792 \\ \hline
Swin\_base\_patch4\_window7\_224 & 0.7560 & 0.7181 & 0.7236 & 0.7677 & 0.2665 & 0.6464 \\ \hline
\end{tabular}
\end{table}

\begin{table}[htb]
\centering
\caption{Performance of different checkpoints on the test set.}
\label{label5}
\begin{tabular}{|l|c|c|}
\hline
\multicolumn{1}{|c|}{Check-points}   & Val Kappa & Test Kappa\\ \hline
V1 - VGG19-Fold0 & 0.8787 &   0.7034         \\ \hline
V2 - VGG16-Fold0 & 0.8592 &   0.7202          \\ \hline
V3 - VGG13-Fold1 & 0.8409 &   0.7045           \\ \hline
V4 - VGG16-Fold1 & 0.8231 &   0.6991             \\ \hline
V5 - VGG19-Fold2 & 0.8933 &   0.7333           \\ \hline
V6 - Efficientnet\_b4-Fold2 & 0.8855 &   0.7184           \\ \hline
V7 - VGG16-Fold3 & 0.9168 &   0.6987           \\ \hline
V8 - VGG19-Fold3 & 0.8632 &   0.7154           \\ \hline
V9 - VGG19-Finetune & 0.8548 & 0.7447  \\ \hline
\end{tabular}
\end{table}

We generated pseudo-labels for each image on the test set based on the baseline model using the classification probabilities. As illustrated in Fig.~\ref{fig1}, we treated prediction results for data with probabilities greater than 0.95 (part 1) as pseudo-labels. In cases where the classification probability was between 0.5 and 0.65 (part 2), pseudo-labels were generated using a hard-voting method.\\

Using all the training sets, we retrained the best-performing four backbones based on the mean of each checkpoint in Table~\ref{label4}. Together with the baseline model, these four checkpoints were hard-voted. Table~\ref{label6} shows the hard voting results for some of the low-confidence data. The pseudo-labels for 14 of the 20 (6 Unsure) part 2 data were generated and added to the training set by repeating them twice.


\begin{table}[htb]
\centering
\caption{Hard voting to produce pseudo labels.}
\label{label6}
\begin{tabular}{|l|c|c|c|c|c|c|}
\hline
\multicolumn{1}{|c|}{Image}   & VGG19 & VGG13 & VGG16 & VGG11 & Efficientnet\_b4 & Pseudo label\\ \hline
952.png & 2 & 1  & 2 & 1 & 1 &    Unsure    \\ \hline
986.png & 2 & 2  & 1 & 2 & 1 &    2     \\ \hline
1220.png& 1 & 2 & 1 & 1 & 2 &     1        \\ \hline
1283.png & 2 & 1 & 1 & 1 & 1 &     1       \\ \hline
901.png & 2 & 1  & 1& 2 & 1 &      Unsure       \\ \hline
897.png & 2 & 2  & 2 & 2 & 2 &        2    \\ \hline
611.png & 1& 1 & 2 & 1 & 1  &     1       \\ \hline
\end{tabular}
\end{table}

Table~\ref{label7} illustrates the effectiveness of our pseudo-label learning method. Following pseudo-label learning with the data from part 1, the Kappa value increased from 0.7447 to 0.7484. As a result of pseudo-label learning using data from part 2, the Kappa value increased from 0.7447 to 0.7513. The results indicate that both parts of the data are essential for improving the classification performance. The baseline model's classification performance was significantly improved with a Kappa value of 0.7589 when both parts of the data were used for pseudo-label learning.\\

In addition to the baseline model VGG19, we also performed pseudo-label learning on VGG16, resulting in a Kappa of 0.7547. Thus, our Kappa improved to 0.7662 after performing the model ensemble on VGG19 and VGG16. \\

This result was used to update the pseudo-labels in the second round. There are 399 images from part 1, and 7 from part 2 (4 pseudo labels, 3 unsure). As a result of the first round of pseudo-label learning, the model was also enhanced. The updated pseudo-labels were used to finetune VGG19, resulting in a kappa value of 0.7803.

\begin{table}[htb]
\centering
\caption{Pseudo-label Ablation study}
\label{label7}
\begin{tabular}{|l|c|c|}
\hline
\multicolumn{1}{|c|}{Method}   & Val Kappa & Test Kappa\\ \hline
VGG19 (Baseline)  &   0.8548  & 0.7447       \\ \hline
\makecell[l]{VGG19-Pseudo-label \\ part 1 }   & 0.9458 &  0.7484         \\ \hline
\makecell[l]{VGG19-Pseudo-label \\ part 2}  & 0.8830 &   0.7513           \\ \hline
\makecell[l]{VGG19-Pseudo-label \\ part 1 + part 2} & 0.8733 &   0.7589           \\ \hline
\end{tabular}
\end{table}



\subsection{Classification results}
The objective of this third task is DR grading. The training dataset groups 611 patients into three grades: label (0) - Normal - (329 images), label (1) - NRDP - (212 images), and finally label (2) - RDP - (70 images). \\

To process this task, we followed the same steps as task 2. Firstly, we performed five-fold cross-validation and selected the eight best-performing checkpoints on the test set. The two most appropriate checkpoints were selected based on the kappa in the test set: V1 - DenseNet121-Fold1 and V2 - Efficientnet b3-Fold3. The baseline model was then fine-tuned using all the training sets: V3 - DenseNet121-Finetune and V4 - Efficientnet\_b3-Finetune. Secondly, we generated pseudo-labels based on the classification results of V3 - DenseNet121-Finetune. In the first round of pseudo-label learning, there were 266 images for part 1, and 20 images for part 2 (15 pseudo labels, 5 unsure). We then performed the first round of pseudo-label learning for DenseNet121 and Efficientnet\_b3. Next, we performed a model ensemble (Kappa = 0.8628) and obtained new pseudo-labels. In the second round of pseudo-label learning, there were 332 images for part 1 and 12 images for part 2 (7 pseudo labels, 5 unsure). After the second round of pseudo-label learning, we performed a model ensemble on DenseNet121 and Efficientnet\_b3 and obtained the final Kappa 0.8761.

\begin{table}[htb]
\centering
\caption{Classification results for different steps}
\label{tab8}
\begin{tabular}{|l|c|c|}
\hline
\multicolumn{1}{|c|}{Chick-point}   & Val Kappa & Test Kappa\\ \hline
V1 - DenseNet121-Fold1 & 0.8275 &  0.8100           \\ \hline
V2 - Efficientnet\_b3-Fold3 & 0.8776  &   0.8069           \\ \hline
V3 - DenseNet121-Finetune & 0.9335 &  0.8370 \\ \hline
V4 - Efficientnet\_b3-Finetune & 0.9728 &  0.8239 \\ \hline
V5 - DenseNet121-First round & 0.9542  & 0.8499 \\\hline
V6 - Efficientnet\_b3-First round & 0.9112 & 0.8545 \\\hline
V7 - DenseNet121-Second round & 0.9674 & 0.8520 \\\hline
V8 - Efficientnet\_b3-Second round  & 0.9558 & 0.8662 \\\hline
\end{tabular}
\end{table}





\section{Conclusion}
In this article, we summarized our participation in the DRAC22 challenge. 
We showed that despite the efficiency of the nnU-Net method in the segmentation task, it does not always give good results, especially when the data set is relatively small. However, the fine-tuning of the V-Net model allowed us to overcome this limitation by obtaining better results for both labels (1) and (3). \\

During the test phase, we noticed many images containing label (3), which was inconsistent with the distribution of the training set. Adding an independent model for the binary classification of the images (either containing label (3) or not) before the segmentation improved our result for this label.\\

For tasks 2 and 3, the pseudo-labeling allowed us to improve our models progressively. Indeed, training baselines, using them to label the test set, and then keeping the labeled images with high confidence allows the model to have more data in the second training round. This iterative process allowed our models to perform better.\\

We have also shown that ensembles of models can generate good performance and allow us to label low-confident data. Indeed, the ensembles can overcome bias and variance from different architectures. Models help each other and cancel each other's errors, resulting in higher accuracy.\\

As for our work in progress, we are combining tasks for better segmentation and classification. We think that using the segmentation results could guide the classifier of task 3. Also, we noticed that when the image quality is poor, this image is still segmented in task 1, giving us segmented regions that should not exist. So we believe that using the models of task 2 before the segmentation could improve the performance.

\bibliographystyle{splncs04}
\bibliography{ref}

\begin{thebibliography}{10}
\providecommand{\url}[1]{\texttt{#1}}
\providecommand{\urlprefix}{URL }
\providecommand{\doi}[1]{https://doi.org/#1}

\bibitem{Atwany}
Atwany, M.Z., Sahyoun, A.H., Yaqub, M.: Deep learning techniques for diabetic
  retinopathy classification: A survey. IEEE Access  \textbf{10},  28642--28655
  (2022). \doi{10.1109/ACCESS.2022.3157632}

\bibitem{00f07def2e1c497d99207b316b574b21}
Dai, L., Wu, L., Li, H., Cai, C., Wu, Q., Kong, H., Liu, R., Wang, X., Hou, X.,
  Liu, Y., Long, X., Wen, Y., Lu, L., Shen, Y., Chen, Y., Shen, D., Yang, X.,
  Zou, H., Sheng, B., Jia, W.: A deep learning system for detecting diabetic
  retinopathy across the disease spectrum. Nature Communications
  \textbf{12}(1) (Dec 2021). \doi{10.1038/s41467-021-23458-5}

\bibitem{deCarlo2015}
De~Carlo, T.E., Romano, A., Waheed, N.K., Duker, J.S.: A review of optical
  coherence tomography angiography (octa). International Journal of Retina and
  Vitreous  \textbf{1}(1), ~5 (Apr 2015). \doi{10.1186/s40942-015-0005-8}

\bibitem{https://doi.org/10.48550/arxiv.1512.03385}
He, K., Zhang, X., Ren, S., Sun, J.: Deep residual learning for image
  recognition (2015). \doi{10.48550/ARXIV.1512.03385}

\bibitem{https://doi.org/10.48550/arxiv.1608.06993}
Huang, G., Liu, Z., van~der Maaten, L., Weinberger, K.Q.: Densely connected
  convolutional networks (2016). \doi{10.48550/ARXIV.1608.06993},
  \url{https://arxiv.org/abs/1608.06993}

\bibitem{https://doi.org/10.48550/arxiv.1809.10486}
Isensee, F., Petersen, J., Klein, A., Zimmerer, D., Jaeger, P.F., Kohl, S.,
  Wasserthal, J., Koehler, G., Norajitra, T., Wirkert, S., Maier-Hein, K.H.:
  nnu-net: Self-adapting framework for u-net-based medical image segmentation
  (2018). \doi{10.48550/ARXIV.1809.10486},
  \url{https://arxiv.org/abs/1809.10486}

\bibitem{lee2013pseudo}
Lee, D.H., et~al.: Pseudo-label: The simple and efficient semi-supervised
  learning method for deep neural networks. In: Workshop on challenges in
  representation learning, ICML. vol.~3, p.~896 (2013)

\bibitem{Yihao}
Li, Y., El~Habib~Daho, M., Conze, P.H., Al~Hajj, H., Bonnin, S., Ren, H.,
  Manivannan, N., Magazzeni, S., Tadayoni, R., Cochener, B., Lamard, M.,
  Quellec, G.: Multimodal information fusion for glaucoma and diabetic
  retinopathy classification. In: Antony, B., Fu, H., Lee, C.S., MacGillivray,
  T., Xu, Y., Zheng, Y. (eds.) Ophthalmic Medical Image Analysis. pp. 53--62.
  Springer International Publishing, Cham (2022)

\bibitem{LIU2022100512}
Liu, R., Wang, X., Wu, Q., Dai, L., Fang, X., Yan, T., Son, J., Tang, S., Li,
  J., Gao, Z., Galdran, A., Poorneshwaran, J., Liu, H., Wang, J., Chen, Y.,
  Porwal, P., {Wei Tan}, G.S., Yang, X., Dai, C., Song, H., Chen, M., Li, H.,
  Jia, W., Shen, D., Sheng, B., Zhang, P.: Deepdrid: Diabetic
  retinopathy—grading and image quality estimation challenge. Patterns
  \textbf{3}(6),  100512 (2022).
  \doi{https://doi.org/10.1016/j.patter.2022.100512},
  \url{https://www.sciencedirect.com/science/article/pii/S2666389922001040}

\bibitem{https://doi.org/10.48550/arxiv.2103.14030}
Liu, Z., Lin, Y., Cao, Y., Hu, H., Wei, Y., Zhang, Z., Lin, S., Guo, B.: Swin
  transformer: Hierarchical vision transformer using shifted windows (2021).
  \doi{10.48550/ARXIV.2103.14030}, \url{https://arxiv.org/abs/2103.14030}

\bibitem{https://doi.org/10.48550/arxiv.2201.03545}
Liu, Z., Mao, H., Wu, C.Y., Feichtenhofer, C., Darrell, T., Xie, S.: A convnet
  for the 2020s (2022). \doi{10.48550/ARXIV.2201.03545},
  \url{https://arxiv.org/abs/2201.03545}

\bibitem{milletari2016vnet}
Milletari, F., Navab, N., Ahmadi, S.A.: V-net: Fully convolutional neural
  networks for volumetric medical image segmentation (2016)

\bibitem{Quellec22}
Quellec, G., Li, Y., Al~Hajj, H., Bonnin, S., Ren, H., Manivannan, N.,
  Magazzeni, S., Tadayoni, R., Conze, P.H., Lamard, M.: 3-d style transfer
  between structure and flow channels in oct angiography. Invest. Ophthalmol.
  Vis. Sci.  \textbf{63}(7),  2989--F0259 (2022).
  \doi{10.1109/ACCESS.2022.3157632}

\bibitem{ExplAIn}
Quellec, G., {Al Hajj}, H., Lamard, M., Conze, P.H., Massin, P., Cochener, B.:
  Explain: Explanatory artificial intelligence for diabetic retinopathy
  diagnosis. Medical Image Analysis  \textbf{72},  102118 (2021).
  \doi{https://doi.org/10.1016/j.media.2021.102118},
  \url{https://www.sciencedirect.com/science/article/pii/S136184152100164X}

\bibitem{Quellec17}
Quellec, G., Charrière, K., Boudi, Y., Cochener, B., Lamard, M.: Deep image
  mining for diabetic retinopathy screening. Medical Image Analysis
  \textbf{39},  178--193 (2017).
  \doi{https://doi.org/10.1016/j.media.2017.04.012},
  \url{https://www.sciencedirect.com/science/article/pii/S136184151730066X}

\bibitem{ronneberger2015unet}
Ronneberger, O., Fischer, P., Brox, T.: U-net: Convolutional networks for
  biomedical image segmentation (2015)

\bibitem{Russell}
Russell, J., Shi, Y., Hinkle, J., Scott, N., Fan, K., Lyu, C., Gregori, G.,
  Rosenfeld, P.: Longitudinal wide-field swept-source oct angiography of
  neovascularization in proliferative diabetic retinopathy after panretinal
  photocoagulation. ophthalmol retina. Retina  \textbf{3}(4),  350--361 (2019).
  \doi{10.1016/j.oret.2018.11.008}

\bibitem{Schaal}
Schaal, K.B., Munk, M.R., Wyssmueller, I., Berger, L.E., Zinkernagel, M.S.,
  Wolf, S.: Vascular abnormalities in diabetic retinopathy assessed with
  swept-source optical coherence tomography angiography widefield imaging.
  Retina  \textbf{39}(1),  79--87 (2019). \doi{10.1097/IAE.0000000000001938}

\bibitem{10.3389/fpubh.2022.971943}
Sheng, B., Chen, X., Li, T., Ma, T., Yang, Y., Bi, L., Zhang, X.: An overview
  of artificial intelligence in diabetic retinopathy and other ocular diseases.
  Frontiers in Public Health  \textbf{10} (2022).
  \doi{10.3389/fpubh.2022.971943},
  \url{https://www.frontiersin.org/articles/10.3389/fpubh.2022.971943}

\bibitem{shrivastava2016training}
Shrivastava, A., Gupta, A., Girshick, R.: Training region-based object
  detectors with online hard example mining. In: Proceedings of the IEEE
  conference on computer vision and pattern recognition. pp. 761--769 (2016)

\bibitem{https://doi.org/10.48550/arxiv.1409.1556}
Simonyan, K., Zisserman, A.: Very deep convolutional networks for large-scale
  image recognition (2014). \doi{10.48550/ARXIV.1409.1556},
  \url{https://arxiv.org/abs/1409.1556}

\bibitem{https://doi.org/10.48550/arxiv.1905.11946}
Tan, M., Le, Q.V.: Efficientnet: Rethinking model scaling for convolutional
  neural networks  (2019). \doi{10.48550/ARXIV.1905.11946},
  \url{https://arxiv.org/abs/1905.11946}

\bibitem{https://doi.org/10.1111/aos.14299}
Tian, M., Wolf, S., Munk, M.R., Schaal, K.B.: Evaluation of different
  swept’source optical coherence tomography angiography (ss-octa) slabs for
  the detection of features of diabetic retinopathy. Acta Ophthalmologica
  \textbf{98}(4),  e416--e420 (2020). \doi{https://doi.org/10.1111/aos.14299},
  \url{https://onlinelibrary.wiley.com/doi/abs/10.1111/aos.14299}

\bibitem{rw2019timm}
Wightman, R.: Pytorch image models.
  \url{https://github.com/rwightman/pytorch-image-models} (2019).
  \doi{10.5281/zenodo.4414861}

\bibitem{Zeghlache}
Zeghlache, R., Conze, P.H., El~Habib~Daho, M., Tadayoni, R., Massin, P.,
  Cochener, B., Quellec, G., Lamard, M.: Detection of diabetic retinopathy
  using longitudinal self supervised learning. In: Antony, B., Fu, H., Lee,
  C.S., MacGillivray, T., Xu, Y., Zheng, Y. (eds.) Ophthalmic Medical Image
  Analysis. pp. 43--52. Springer International Publishing, Cham (2022)

\bibitem{QIMS21249}
Zhang, Q., Rezaei, K.A., Saraf, S.S., Chu, Z., Wang, F., Wang, R.K.: Ultra-wide
  optical coherence tomography angiography in diabetic retinopathy.
  Quantitative Imaging in Medicine and Surgery  \textbf{8}(8) (2018),
  \url{https://qims.amegroups.com/article/view/21249}

\end{thebibliography}
\end{document}